\documentclass[11pt]{article}
\usepackage{moriond,epsfig}

\def\be{\begin{equation}}
\def\ee{\end{equation}}
\def\bea{\begin{eqnarray}}
\def\eea{\end{eqnarray}}

\newcommand{\MET}{$\not\!\!E_T$}
\def\pt{$p_T$}
\def\Aboson{$A$}
\def\Hboson{$H$}
\def\hboson{$h$}
\def\bbbar{$b\bar{b}$}

\def\tanb{$\tan{\beta}$}

\def\btag{$b$}
\begin{document}
\vspace*{4cm}
\title{Searches for Non-SM Higgs at the Tevatron}

\author{Makoto Tomoto \\ (for D\O\ and CDF collaborations)}

\address{Fermi National Accelerator Laboratory, \\
P.O. Box 500, Batavia, IL, 605100-0500, USA}

\maketitle\abstracts{
We present the non-standard model Higgs boson searches at both D\O\ and CDF experiments.
We concentrate on three topics, reported by D\O\ and CDF recently: the 
searches for the minimal supersymmetric Higgs boson using $p\bar{p} \rightarrow hb\bar{b} \rightarrow 
b\bar{b}b\bar{b}$ and $p\bar{p} \rightarrow hX \rightarrow \tau\tau X$, and for long-lived doubly 
charged Higgs boson.
}

\section{Introduction}
Tevatron is currently the only place where we can search for not only 
the standard model (SM) Higgs boson but also non-SM Higgs bosons.
CDF and D\O\ have started searching for the exotic Higgs bosons 
including supersymmetric Higgs bosons and doubly charged 
Higgs bosons.
D\O\ recently updated the results of the minimal supersymmetric standard model 
(MSSM) Higgs boson search using $hb\bar{b} \rightarrow 
b\bar{b}b\bar{b}$ final state. CDF updated its search for 
MSSM Higgs bosons using  $hX \rightarrow \tau \tau X$
final state, and long-lived doubly charged Higgs bosons.

\section{Search for MSSM Higgs bosons}
In two-Higgs-doublet models of electroweak symmetry breaking, such
as the MSSM~\cite{2HDM}, there are five physical Higgs bosons:
two neutral $CP$-even scalars, $h$ and
$H$; a neutral $CP$-odd state, $A$; and two charged states, $H^\pm$. The ratio of the vacuum
expectation values of the two Higgs fields is defined as $\tan{\beta}$ =
$v_2/v_1$, where $v_2$ and $v_1$ refer to the Higgs fields that couple to
the up-type and down-type fermions, respectively.
At tree level, the coupling of the $A$ boson to down-type quarks,
such as the $b$ quark,
is enhanced by a factor of $\tan{\beta}$ relative to the standard model (SM),
and the production
cross section is therefore enhanced by $\tan^{2}\beta$~\cite{MSSM_Higgs}.
The Tevatron is currently sensitive to \tanb\ in the range $\tan{\beta}>$ 50.
In this region of \tanb, the $A$ boson is nearly degenerate in
mass with either the \hboson\ or the \Hboson\ boson, and their
widths are small compared to the di-jet mass resolution.
Consequently, we cannot distinguish between the \hboson/\Hboson\
and the \Aboson, and the total cross section for signal is
assumed to be twice that of the $A$ boson.
LEP experiments have excluded at the 95\% C.L. a light Higgs boson with mass
$m_h$ $\le$ 92.9~GeV~\cite{leplimit}.
The major decay modes of $h$ 
are a $b$ quark pair production(90\%) and $\tau$ lepton pair production(8\%).
The $p\bar{p} \rightarrow hb\bar{b}$ with $h\rightarrow b\bar{b}$ 
or $h \rightarrow \tau\tau$ decay are considered the most promising 
channels at Tevatron. 

\subsection{Search for MSSM Higgs bosons in $hbb$ channel at D\O }

The analysis is based on 260~$\mbox{pb}^{-1}$ of data collected using the D\O\ detector from November 2002 to
June 2004.
The dedicated trigger is designed for maximizing signal acceptance 
while providing reasonable trigger rate.
Events with up to five jets are selected.
Depending on the hypothesized Higgs boson
mass, the jet \pt\ selections are chosen to optimize the expected 
signal significance.
Jets containing \btag\ quarks are identified using a secondary
vertex (SV) tagging algorithm. 
The \btag\ tagging efficiency is $\approx$~55\% for central
\btag-jets of \pt$>$35~GeV with a light quark (or gluon) tag rate
of about 1\%.

The signal $bh$ events, with $h\rightarrow b\bar{b}$, were 
generated for Higgs boson masses from 90 to 150~GeV using 
{\footnotesize PYTHIA} event generator~\cite{pythia}.
The \pt\ and rapidity spectra of the Higgs bosons from 
{\footnotesize PYTHIA} were adjusted to those from NLO calculation~\cite{5fns}.
The multijet production is the major source of
background and is  determined from data. 
As a cross-check, we
also compared the data with {\sc alpgen}~\cite{alpgen} event generator.
All other backgrounds are expected
to be small and are simulated with {\sc pythia}.

\begin{figure}[!htpb]
\begin{center}
\psfig{figure=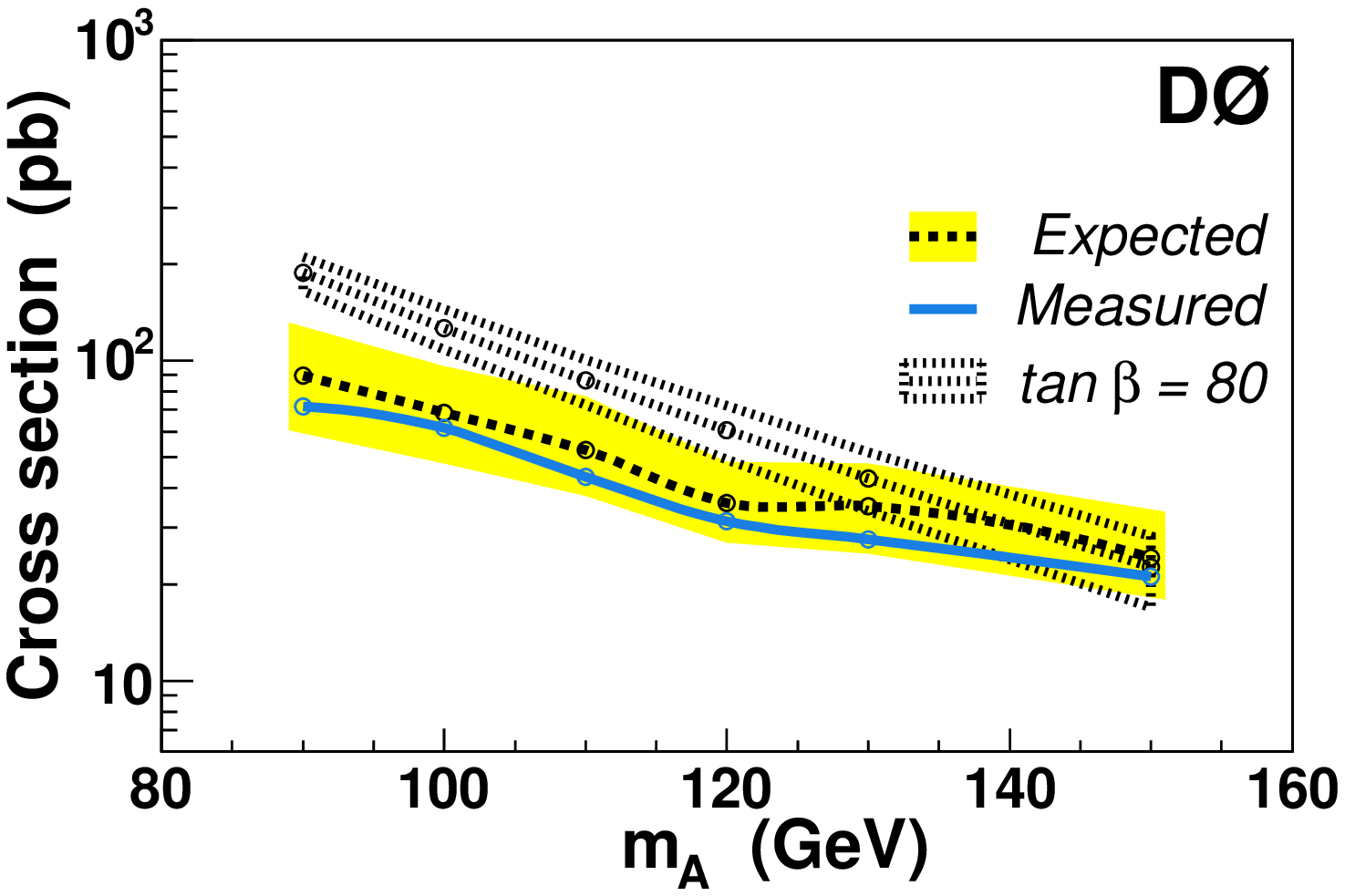,height=1.4in}
\psfig{figure=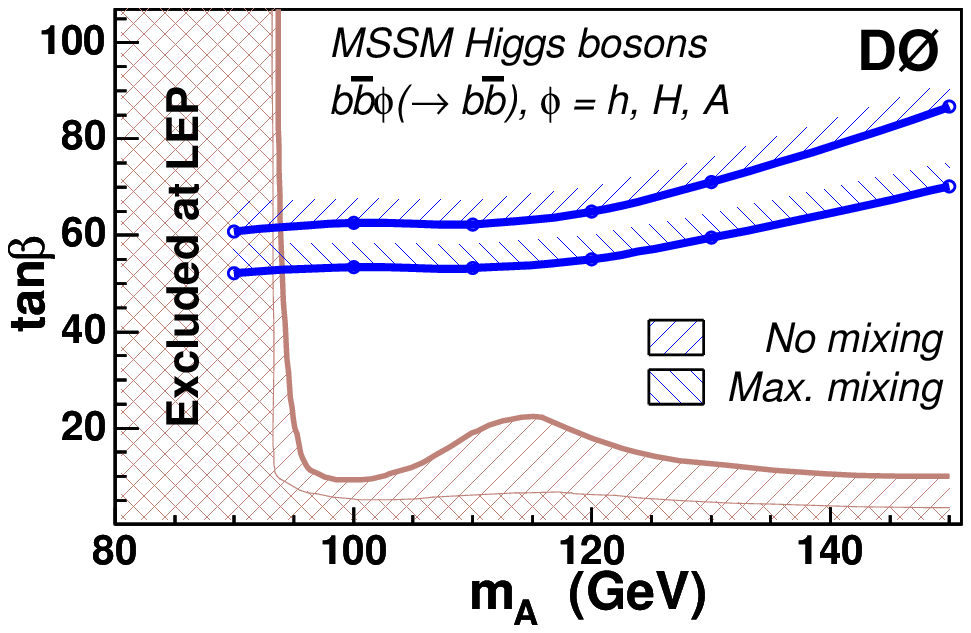,height=1.4in}
\end{center}
\caption{The upper limits on the signal cross section as a function of Higgs boson mass~(left) 
and the 95\% C.L. upper limit on $\tan{\beta}$ as a function of Higgs boson mass~(right).} \label{fig:mh_limit}
\end{figure}

There are two main categories of multijet background. One contains
genuine heavy-flavor jets, while the other has only light-quark
or gluon jets that are mistakenly tagged as \btag-quark jets, or
correspond to gluons that branch into nearly collinear \bbbar\
pairs. Using the selected data sample, before the application of
\btag-tagging requirements, the probability to \btag-tag a jet (``mis-tag'' function) is
measured as a function of its \pt\ and $|\eta |$. 
The mis-tag function is corrected by subtracting heavy-flavor contributions.
This mis-tag function is used to estimate
the mis-tagged background, by applying it to every jet
reconstructed in the full data sample.
The triple $b$-tagged data are compared with this background expectation 
and no signal evidence is found.
Figure~\ref{fig:mh_limit} shows the expected MSSM Higgs boson
production cross section as a function of $m_A$ for \tanb\ = 80. 
The MSSM cross
section shown in Figure~\ref{fig:mh_limit} corresponds to no mixing in
the scalar top quark sector~\cite{tev}, or $X_{t}$ = 0, where
${X_{t} = A_{t} - \mu\cot\beta}$, $A_{t}$ is the tri-linear
coupling, and the Higgsino mass parameter $\mu = -0.2$~TeV. We also
interpret our results in the ``maximal mixing'' scenario with $X_{t}
= \sqrt6 \times M_{{SUSY}}$, where $M_{{SUSY}}$ is the
mass scale of supersymmetric particles, taken to be 1~TeV.
Results for both scenarios of the MSSM are shown in
Figure~\ref{fig:mh_limit} as limits in the \tanb\ versus $m_A$ plane.
The present D\O\ analysis excludes a
significant portion of the parameter space, down to \tanb = 50,
depending on $m_A$ and the MSSM scenario assumed.

\subsection{Search for MSSM Higgs bosons in $hX \rightarrow \tau \tau X$ channel at CDF}

Using the data collected by the CDF detector from March 2002 to January 2004, corresponding to
an integrated luminosity of about 200 $\mbox{pb}^{-1}$, MSSM Higgs searches for 
the $hX \rightarrow \tau \tau X$ is presented.
The signal consists of a tau pair in which one of the taus decays to hadrons and a 
neutrino ($\tau_h$) while the other decays to an electron~($\tau_e$) or a muon~($\tau_\mu$) and two neutrinos.
The data are collected with a set of dedicated $\tau$-triggers which 
select a lepton ($e$ or $\mu$) candidate with $p_T>$ 8~GeV and an isolated track from $\tau_h$ decay.
The $\tau$, which decays to hadrons, are reconstructed by charged tracks and neutral pions inside 
special $\tau$ narrow cone. The invariant mass of $\tau_h$ and 
the track multiplicity of the $\tau_h$ is required to be less than 1.8~GeV and equal to 
1 or 3, respectively.
The further selections are done on the event topology using 
missing transverse energy (\MET), $p_T$ of $\tau_e$ or $\tau_\mu$ and $\tau_h$. 
An example is the scalar sum $\hat{H_T} = |p_T^{vis}(\tau_1)|+|p_T^{vis}(\tau_2)|+$\MET, 
where $p_T^{vis}(\tau)$ is visible transverse momentum of the tau decay products not including neutrinos.
Requiring $\hat{H_T}>$50~GeV leads to significant background reduction with small signal loss.

\begin{figure}[!htpb]
\begin{center}
\psfig{figure=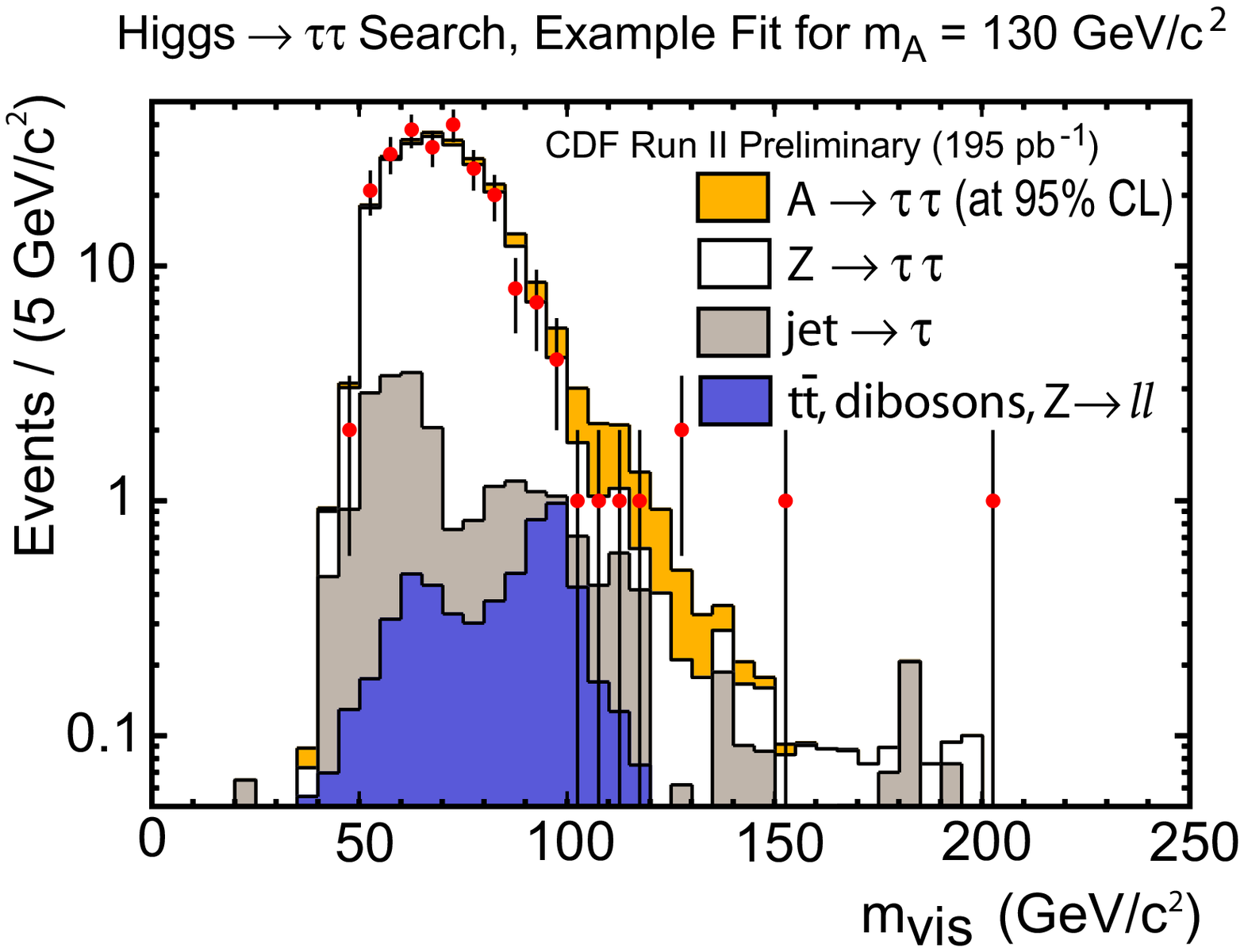,height=1.4in}
\psfig{figure=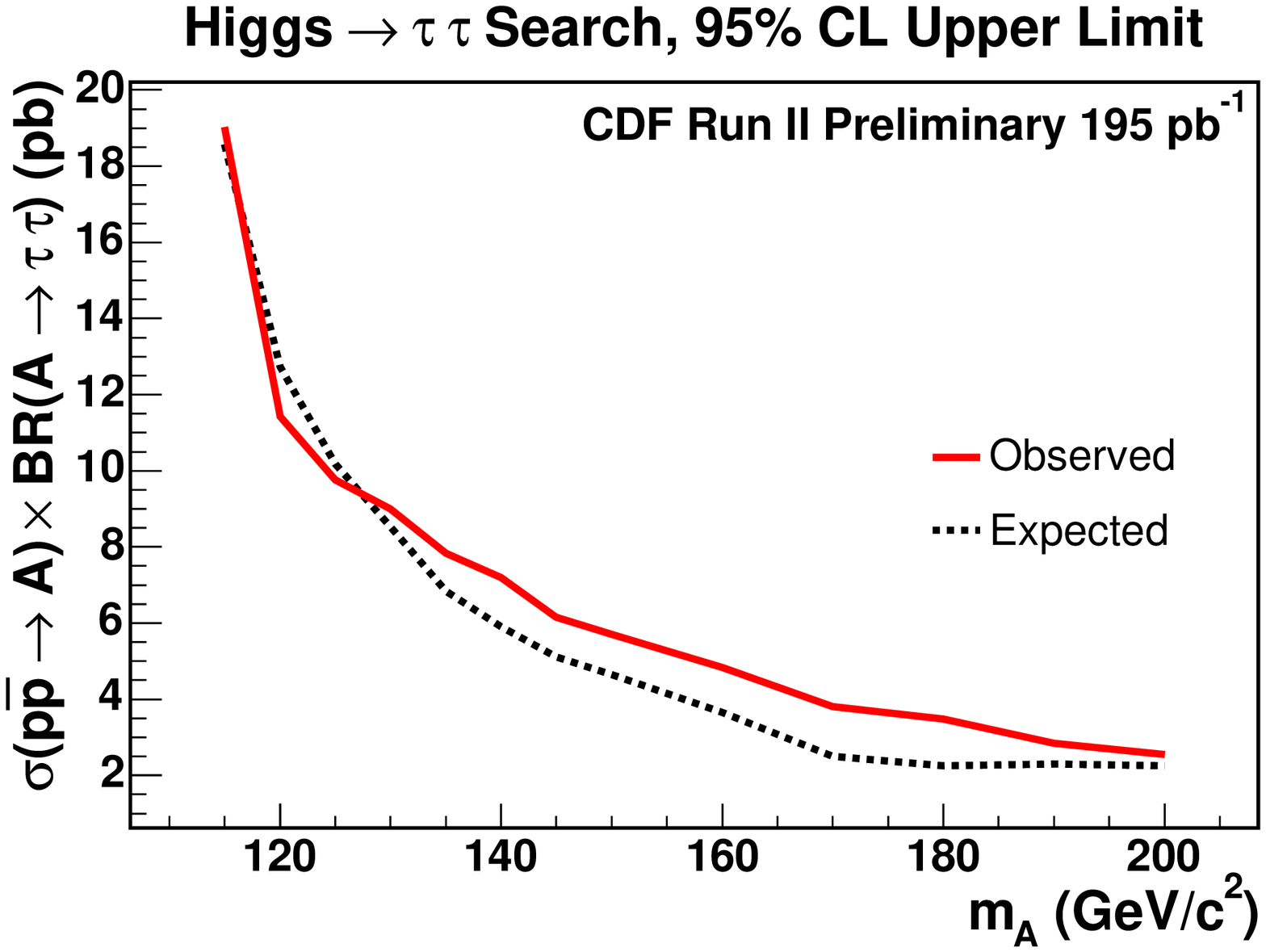,height=1.4in}
\end{center}
\caption{Observed $m_{vis}$($l$,$\tau_h^{vis}$, \MET) distributions for various background~(left) and 
limits on the production product 
$\sigma(p\bar{p} \rightarrow h) \times B(A \rightarrow \tau\tau)$~(right).} \label{fig:mass_fit}
\end{figure}

The signal processes $gg\rightarrow h$ and $bb\rightarrow h$ is simulated using
{\footnotesize PYTHIA} event generator. We generate Higgs masses between 115~GeV and 
200~GeV for $\tan{\beta}=$30. The backgrounds from $Z\rightarrow l^{+}l^{-}$, 
($l$=$e$,$\mu$,or $\tau$), di-boson production, and $t\bar{t}$ production 
are estimated using Monte Carlo samples. The Backgrounds from jet to $\tau$ misidentification 
are estimated using a fake rate function obtained from independent jet samples, as a function 
of the jet energy, psuedorapidity, and track multiplicity. 
The biggest source of background is $Z\rightarrow \tau\tau$ events,
which can only be distinguished from the Higgs signal by the di-$\tau$  mass.
We define a mass-like discriminating variable $m_{vis}$($l$,$\tau_h^{vis}$, \MET).
It is constructed using the four-momentum of the lepton ($e$ or $\mu$), the 
four-momentum of the visible decay products of the $\tau_h$($\tau_h^{vis}$)
and \MET\ (also treated as a four vector). After combining the $\tau_e\tau_h$ 
and $\tau_h\tau_\mu$ channels, we perform a binned likelihood fit of the $m_{vis}$($l$,$\tau_h^{vis}$, \MET)
distribution to extract a Higgs signal. The uncertainties in background estimation are incorporated 
in the imposed Gaussian constraints. Several signal templates were generated to cover the studied mass region.
 Figure~\ref{fig:mass_fit} shows the $m_{vis}$($l$,$\tau_h^{vis}$, \MET) distribution for data and various backgrounds. The limits on Higgs production cross section times branching fraction at 95 \% C.L. are presented in Figure~\ref{fig:mass_fit}. 

\section{Search for long-lived doubly charged Higgs bosons at CDF}

A number of models like the left-right symmetric model requires a Higgs triplet containing 
a doubly-charge Higgs boson~($H^{\pm\pm}$), which could be light in the minimal supersymmetric left-right model.
The dominant production mode at the Tevatron is expected to be 
$p\bar{p} \rightarrow \gamma^{*}/Z + X \rightarrow H^{++} H^{--}$.
The partial width in the leptonic decay mode is proportional to the coupling to the lepton and Higgs mass.
D\O\ and CDF published the $H^{\pm\pm}$ mass limits from direct searches in the di-lepton decay channels for 
the short lived $H^{\pm\pm}$~\cite{dhiggsSL_D0}$^,\ $\cite{dhiggsSL_CDF}. We present here the case where $H^{\pm\pm}$ boson is long-lived~($c\tau>$3~m), 
resulting in the $H^{\pm\pm}$ boson decaying outside the detector. DELPHI experiment~\cite{delphi} excludes $m_{H^{\pm\pm}}<$99.6~GeV
(99.3~GeV) at the 95\% C.L. for $H^{\pm\pm}$ bosons with coupling to left-(right-) handed leptons.

The analysis is based on 292 $\mbox{pb}^{-1}$ of data collected using the CDF Run~II detector.
We use an inclusive muon trigger requiring a track with  $p_T>$ 18~GeV, and a matching with a track in the muon chamber.
We select two tracks, each with $p_T>$20~GeV, and at least one of the tracks to have 
a matching muon. Since the charge of tracks 
collected by drift chamber is proportional to the ionization deposited by 
the particle per unit length($dE/dx$), $H^{\pm\pm}$ boson deposits 
four times more  $dE/dx$ in the drift chamber than a single charge track like a $\mu$. 
The $H^{\pm\pm}$ boson is modeled by quadrupling the $dE/dx$ of the cosmic ray muons.
The $dE/dx$ from the low momentum protons are used as a controlled sample.
We calculate the geometric and kinematic acceptance for a pair of $H^{\pm\pm}$ boson using the 
{\footnotesize PYTHIA} generator.

\begin{figure}[!htpb]
\begin{center}
\psfig{figure=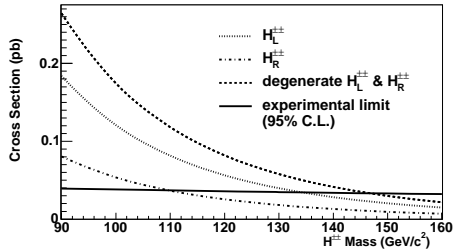,height=1.4in}
\end{center}
\caption{The comparison of the experimental cross section upper limit with the theoretical cross section
 for pair production of $H^{\pm}$ bosons.} \label{fig:dhiggs_limit}
\end{figure}

Since no data remained after selecting large $dE/dx$, we set 95\% C.L. upper limits on the $H^{\pm}$ pair production cross section, as shown Figure~\ref{fig:dhiggs_limit}.
The theoretical cross sections are computed separately for $H^{\pm}$ bosons that couple to left- and right-handed particles($H^{\pm}_L$ and $H^{\pm}_R$). 
We exclude the long-lived $H^{\pm}_L$ and $H^{\pm}_R$ bosons below 
a mass of 133~GeV and 109~GeV, respectively. 
When the two states are degenerate in mass, we exclude 
$H^{\pm}<$146~GeV at the 95~\% C.L.
\section{Conclusion}
We reported on searches for several types of exotic Higgs bosons.
We have excluded significant area of theoretical parameters.
D\O\ and CDF are expected to collect  more data and employ 
advanced analysis techniques, therefore, 
will continue exploring the Higgs domain.



\section*{References}
\begin{thebibliography}{99}
\bibitem{2HDM}
H.~P.~Nilles,
Phys.\ Rept.\  {\bf 110}, 1 (1984);
H.~E.~Haber and G.~L.~Kane,
Phys.\ Rept.\  {\bf 117}, 75 (1985).

\bibitem{MSSM_Higgs}
J.~F.~Gunion, H.~E.~Haber, G.~L.~Kane, and S.~Dawson, ``The Higgs
Hunter's Guide,'' Addison-Wesley, 1990.

\bibitem{leplimit}
The LEP Working Group for Higgs Boson Searches, LHWG-Note 2004-01.

\bibitem{pythia}
T.~Sj{\"o}strand {\sl et al.},
Comp.\ Phys.\ Comm.\  {\bf 135}, 238 (2001).

\bibitem{5fns}
J.~Campbell, R.~K.~Ellis, F.~Maltoni, and S.~Willenbrock,
Phys.\ Rev.\ D~{\bf 67}, 095002 (2003).

\bibitem{alpgen}
M.~L.~Mangano {\sl et al.},
JHEP {\bf 0307}, 001 (2003).

\bibitem{tev}
M.~Carena, S.~Mrenna, and C.~E.~M.~Wagner,
Phys.\ Rev.\ D~{\bf 60}, 075010 (1999).

\bibitem{dhiggsSL_D0}
 D\O\ Collaboration,
V.M.~Abazov {\it et al.},  Phys. Rev. Lett. \bf 93\rm, 141801 (2004).

\bibitem{dhiggsSL_CDF}
 CDF Collaboration,
D.~Acosta {\it et al.},  Phys. Rev. Lett. \bf 93\rm, 221802 (2004).

\bibitem{delphi}
 DELPHI Collaboration, J. Abdallah {\it et al.},
 Phys.~Lett.~B \bf 552\rm, 127 (2003).


\end{thebibliography}
\end{document}